\documentclass[aps, prd, twocolumn, showpacs, amsmath, amssymb, preprintnumbers]{revtex4}
\usepackage{graphicx}
\usepackage{dcolumn}
\usepackage{bm}
\usepackage{mathbbol}
\newcommand{\be}{\begin{equation}}
\newcommand{\ee}{\end{equation}}
\newcommand{\bea}{\begin{eqnarray}}
\newcommand{\eea}{\end{eqnarray}}
\newcommand{\ba}{\begin{array}}
\newcommand{\ea}{\end{array}}

\begin{document}

\title{Viscosity and jet quenching from a holographic model  }
\vskip.3in
\author{ Yi-hong Gao$^a$, Wei-shui Xu$^a$ and Ding-fang Zeng$^b$ }
\affiliation{${}^a${Institute of Theoretical Physics}\\
{P.O. Box 2735,~Beijing 100080, P.~R.~China}\\
${}^b${College of Applied Science, Beijing University of
Technology}\\{Beijing 100022, P.~R.~China}}

\date{June 28, 2007}

\begin{abstract} We consider the
backreaction of the fundamental flavor degrees of freedom on the
AdS$_5$-Schwarz background, and calculate their contributions to the
shear viscosity and jet-quenching parameter of the thermal
quark-gluon plasma.
\end{abstract}


\maketitle

\noindent{\bf Introduction:} From the RHIC experiments, the quark
gluon plasma(QGP) produced in the relativistic heavy ion
collision(RHIC) is strongly coupled in a wide temperature region.
Thus, it is difficult to study this physics by the finite
temperature perturbative QCD. How can one gain insight into the
physics nonperturbatively? We must thank the anti-de
Sitter/conformal field theory(ads/cft) correspondence
\cite{Maldacena:1997re}. In this holographic dual correspondence,
the boundary $\mathcal{N}=4$ supersymmetric Yang-Mills (SYM) theory
in the large $N_c$ and 't Hooft coupling constant $\lambda$ limit is
dual to the ten dimensional supergravity. Actually, at high
temperature, the confinement and chiral condensate in the thermal
QGP of QCD will disappear. Therefore, the $\mathcal{N}=4$ SYM QGP
can share some common dynamical properties with the thermal QCD
quark gluon plasma. Accordingly, one can understand some features of
the QGP produced in the RHIC through studying the $\mathcal{N}=4$
SYM QGP.

There are many properties of QGP in RHIC, such as the shear
viscosity $\eta$, and the energy loss of partons moving through the
QGP, described by the parameter $\hat{q}$ \cite{Baier:1996sk} can be
studied by ads/cft. For example, in \cite{Policastro:2001yc,
Kovtun:2004de, Kovtun:2003wp, Policastro:2002se, Buchel:2003tz,
Buchel:2004di}, the shear viscosity is studied by this method, with
the result that there exists a universal bound for the ratio between
the shear viscosity and entropy density, $ \eta/s \geq 1/4\pi $. For
the $\mathcal{N}=4$ SYM, this bound is saturated. In order to
investigate the QCD-like theories, we need introduce fundamental
flavors and break some supersymmetries. In a recent paper
\cite{Mateos:2006yd}, the authors considered the flavor branes'
backreaction on the background, and compute the corresponding
corrections(proportional to ${N_f/N_c}$ at the leading order) to the
shear viscosity.

For the jet-quenching parameter, reference \cite{Liu:2006ug} gives a
nonperturbative definition through considering a light-like wilson
loop in the $\mathcal{N}=4$ SYM. The energy loss phenomenon of
parton moving through the QGP is also studied by many other works
\cite{Herzog:2006gh, Buchel:2006bv, Armesto:2006zv,
Vazquez-Poritz:2006ba, Gubser:2006nz, Argyres:2006yz, Gao:2006uf,
Liu:2006he, Bertoldi:2007sf, Cotrone:2007qa}, \footnote{More
references can see the paper \cite{Liu:2006he}}. In
\cite{Vazquez-Poritz:2006ba}, the author consider a marginal
deformed $\mathcal{N}=4$ SYM theory, and find that the jet quenching
parameter is dependent on the probing brane's position in the
internal manifold. And in \cite{Bertoldi:2007sf} based on the work
\cite{Casero:2006pt}, some investigations about the contribution of
fundamental flavors to the jet-quenching parameter are considered in
a conformal window.

In this paper, we attempt to study the contribution of fundamental
flavors to the shear viscosity and jet quenching parameter. In
\cite{Karch:2006pv, Andreev:2006vy}, using the so called AdS/QCD
method, the authors constructed some holographic QCD-like models,
which produce the linear confinement behavior of QCD theory. The
five-dimensional gravity background of the holographic model in
\cite{Karch:2006pv} is the deformed AdS$_5$ geometry with a dilaton
field.  While finite temperature generalization of this model is
studied in \cite{DaRold:2005zs}. If only the quadratic terms like
$F^2$ in the effective action are considered, this model is
phenomenologically equivalent to the one constructed in
\cite{Andreev:2006vy}, where the background is also AdS$_5$-like,
but with a constant dilaton and a different warp factor. Based on
these AdS/QCD models, the author of \cite{Herzog:2006ra} gives a
holographic prediction about the critical temperature of the
confinement/deconfinement phase transition.

If we consider a black hole generalization of the holographic models
\cite{Karch:2006pv, Andreev:2006vy}, then the boundary theory will
be a finite temperature gauge theory. Thus we can use the
gauge/gravity correspondence to investigate the properties of this
thermal QGP. Motivated by \cite{Karch:2006pv, DaRold:2005zs,
Andreev:2006vy}, we take the black hole background as follows \bea
&&ds^2=\frac{R^2}{z^2}h(z)(-f(z)dt^2+d\vec{x}^2
+f(z)^{-1}dz^2),\nonumber\\
&&h(z)=e^{\varepsilon z^2}, ~~f(z)=1-\frac{z^4}{z_T^4},
\label{background}\eea where the parameter $\varepsilon$ is chosen
to be in the region $0\leq \varepsilon\leq c$, $c\approx 0.45$
\textrm{Gev$^2$}, and $R^4=4\pi g_sN_c\alpha'^2$ \footnote{ We
suppose this geometry has an origins of string theory.}. The Hawking
temperature and entropy density of this background are
$T'=Te^{\varepsilon z_T^2}$,
$s=\frac{R^3}{4z_T^3G_5}e^{\frac{3}{2}\varepsilon z_T^2}$ with the
$G_5=G_{10}/\pi^3R^5$, $16\pi G_{10}=(2\pi)^7g_s^2\alpha'^4$ and
$T=1/\pi z_T$. If $\varepsilon=0$, then the geometry
(\ref{background}) will reduce to the AdS$_5$-Schwarz metric, which
means that the boundary theory is the thermal $\mathcal{N}=4$ SYM.
However, if $\varepsilon=c$, it leads to the black hole
generalization in \cite{Andreev:2006vy}, and then we get the thermal
QGP of this holographic model. Thus, the background
(\ref{background}) can connect the AdS$_5$-Schwarz metric to the
black hole generalization in \cite{Andreev:2006vy} through varying
the parameter $\varepsilon$. This implies that there may exist a RG
flow \footnote{ The background (\ref{background}) doesn't satisfy
the Ricci flow equation under the variable parameter $\varepsilon$
with the boundary conditions, the black hole generalization in
\cite{Andreev:2006vy} and the AdS$_5$-Schwarz geometry.} between the
$\mathcal{N}=4$ SYM and the QCD-like model as conjectured in
\cite{Liu:2006he}. From the ten-dimensional string theory analysis,
we know that, after introducing the flavor branes into the
AdS$_5$-Schwarz geometry, their backreaction will deform this
geometry. So the shear viscosity will obtain corrections from these
flavor branes, as investigated by \cite{Mateos:2006yd}. By this
motivation, in order to obtain the contributions of the fundamental
flavors to some hydrodynamic properties of the dual non-Abelian
plasma, in the following we will assume that the main origin of the
deformation of the AdS$_5$-Schwarz geometry is the backreaction of
fundamental flavors, and ignore the contributions of the other
degrees of freedom. Roughly, by this assumption, we think that we
can get some relations between the fundamental flavors and the shear
viscosity and jet-quenching parameter. Thus, if the ratio
$N_f/N_c\rightarrow 0$, then the parameter $\varepsilon$ vanishes.
This also means that the background will reduce to the
AdS$_5$-Schwarz geometry. Therefore, we suppose that the varying
parameter $\varepsilon$ is a function of the ratio $N_f/N_c$. And so
we let this parameter be equal to \be
\varepsilon=\sum_{n=1}a_n(N_f/N_c)^n. \label{A}\ee

Thus, using the AdS/QCD method, the shear viscosity and
jet-quenching parameter calculated in the background
(\ref{background}) will contain contributions from the fundamental
flavors. In the latter part of this paper, we will calculate these
quantities of the QGP following the routine of
\cite{Policastro:2001yc, Kovtun:2004de, Buchel:2003tz,
Kovtun:2003wp} and \cite{Liu:2006ug} in the background
(\ref{background}). After some analysis, we find that, compared with
the result of the thermal $\mathcal{N}=4$ SYM, the correction is
proportional to $N_f/N_c$. This result can deepen our understanding
on the contribution of fundamental flavors on the shear viscosity
and jet quenching parameter.

Finally, we will give some discussions and conclusions about our
results.

\noindent{\bf Shear viscosity:} For a generic gravity background \be
ds^2=G_{00}(r)dx_0^2+G_{xx}(r)d\vec{x}^2+G_{rr}(r)dr^2+\cdots,\label{generic}\ee
the coordinates $(x_0, \vec{x})$ denote the four-dimensional
spacetime of the holographic gauge theory, $r$ is the transverse
direction, and the other part in the metric isn't relevant to the
shear perturbation. From the \cite{Policastro:2002se,
Buchel:2003tz}, the diffusion coefficient $D$ of the QGP is \be
D=\frac{\sqrt{-G}}{\sqrt{-G_{00}G_{rr}}}
\Bigg\vert_{r=r_0}\int_{r_0}^\infty
dr\frac{-G_{00}G_{rr}}{G_{xx}\sqrt{-G}},\ee where the $r_0$ is the
horizon of background $(\ref{generic})$.

Thus, for the metric (\ref{background}), the corresponding diffusion
coefficient of the QGP is \bea
&&D=\frac{R^3}{z_T^3}h^{3/2}\Bigg\vert_{z=z_T}\int^{z_T}_0dz\frac{z^3}{R^3h^{3/2}}\nonumber\\
&&~~=\frac{1}{9\varepsilon^2z_T^3}(2e^{\frac{3}{2}\varepsilon
z_T^2}-3\varepsilon z_T^2-2).\eea Since the parameter $\varepsilon$
is very small \footnote{If the parameter $\varepsilon$ reaches the
maximum value $c$, and assuming $T\approx 300$ \textrm{Mev}, then
using $z_T=1/\pi T$, so $3\varepsilon z_T^2/2$ is still smaller than
one.}, we can expand the above equation to the first order of the
constant $\varepsilon$ as  \be D=\frac{1}{4\pi
T}+\varepsilon\frac{1}{8\pi^3T^3}, \ee where we used the temperature
$T$ of AdS$_5$-Schwarz black hole. Using the relation between the
diffusion constant and the shear viscosity $D=\eta/T's$, then we can
get the ration between the shear viscosity $\eta$ and the entropy
density $s$ as\be
\frac{\eta}{s}=\frac{1}{4\pi}+\varepsilon\frac{3}{8\pi^3T^2}.\label{shear}\ee
Since $\varepsilon>0$, the result in the equation (\ref{shear})
obviously satisfies the bound condition $\eta/s\geq 1/4\pi$. The
background (\ref{background}) doesn't satisfy the conditions
proposed in \cite{Buchel:2003tz}, hence this ratio can't saturate
the bound as same with the AdS$_5$-Schwarz case. If letting
$\varepsilon=0$, it will reduce to the same value for
$\mathcal{N}=4$ SYM in \cite{Policastro:2001yc}. Hence the second
term of the equation (\ref{shear}) comes from the contributions of
the fundamental flavors. For the metric (\ref{background}), the
entropy density to the leading order $\varepsilon$ is $s=
\frac{\pi^2N_c^2}{2}T^3+\frac{3}{4}\varepsilon N_c^2T$. Thus, from
equation (\ref{shear}), we have \be \eta=\frac{\pi
N_c^2}{8}T^3+\varepsilon\frac{3N_c^2}{8\pi}T. \ee

As the arguments in the introduction, we have obtained the relation
between the parameter $\varepsilon$ and the comparison $N_f/N_c$.
Since the $N_f/N_c$ is smaller than one, we can omit the high order
terms of the $N_f/N_c$. Therefore, the $\varepsilon$ can be
approximate to $a_1N_f/N_c$. Roughly, in the QCD-like side, we can
choose $N_c=3$ and $N_f=1$, which means $N_f/N_c=1/3$. And now the
$\varepsilon$ is equal to the constant $c=0.45$ \textrm{Gev$^2$}.
Then we can fix the constant $a_1=3c$, and the parameter
$\varepsilon=3cN_f/N_c$. So the entropy density and shear viscosity
read \bea s=\frac{\pi^2N_c^2}{2}T^3+\frac{9c}{4}N_cN_fT,
\nonumber\\
\eta=\frac{\pi N_c^2}{8}T^3+\frac{9c}{8\pi}N_cN_fT.\eea And the
ratio between the $\eta$ and $s$ is \be
\frac{\eta}{s}=\frac{1}{4\pi}+\frac{N_f}{N_c}\frac{9c}{8\pi^3T^2}.\ee

Thus, we find that the corrections of the entropy density and  shear
viscosity compared with the results in the $\mathcal{N}=4$ SYM  are
all proportional to the $N_fN_c$, which is similar to the case in
\cite{Mateos:2006yd} for a fixed 't Hooft coupling constant. But the
temperature dependence is different, here is proportional to $T$,
however, in \cite{Mateos:2006yd}, it is proportional to $T^3$. And
for the ratio $\eta/s$, through the analysis of the ten-dimensional
string theory, the authors in \cite{Mateos:2006yd} find that, to the
leading order $N_f/N_c$, the ratio $\eta/s$ does still saturate the
bound condition as same as the $\mathcal{N}=4$ SYM. However, from
the more holographic QCD-like model in the above, we find our result
doesn't saturate the bound, and the correction is proportional to
the parameter $N_f/N_c$. Or in other words, if omitting the high
order terms correction about $N_f/N_c$, the ratio $\eta/s$ is linear
dependent on the $N_f/N_c$. so with varying $N_f/N_c$ from $0$ to
$1/3$, the $\mathcal{N}=4$ SYM will flow to the holographic QCD-like
model in \cite{Karch:2006pv, Andreev:2006vy}.  And another
interesting feature is this correction being temperature dependent
on $T^{-2}$. If the temperature becomes large, then the correction
will become smaller.

In the holographic model \cite{Karch:2006pv, Andreev:2006vy}, i.e
$\varepsilon=c$, or as above choosing $N_c=3$ and $N_f=1$, we get
$s=\frac{9\pi^2}{2}T^3+\frac{27c}{4}T$, and  \bea
\eta= \frac{9\pi }{8}T^3+\frac{27c}{8\pi}T,\nonumber\\
\frac{\eta}{s}=\frac{1}{4\pi}+\frac{3c}{8\pi^3T^2}.\eea Using these
equations, we can compare our results with the RHIC data. If
choosing the temperature $T=300$ \textrm{Mev}, then the shear
viscosity and the ratio $\eta/s$ are \be \eta=0.24~\textrm{Gev$^3$},
~~~\frac{\eta}{s}=0.14. \label{result}\ee From the RHIC data, the
shear viscosity satisfies $\eta<0.35$ \textrm{Gev$^3$}, and
$0\leq\eta/s\ll 1$. So in our results, the shear viscosity and the
ratio $\eta/s$ are all in the region of the RHIC data.

\noindent{\bf Jet quenching parameter:} The jet-quenching parameter
of the holographic model \cite{Karch:2006pv, Andreev:2006vy} was
investigated in \cite{Nakano:2006js}, \footnote{For the drag force
calculation, one also can see the paper \cite{Nakano:2006js}.},
however, we still give some calculations in order to conveniently
discuss about the corrections to the jet-quenching parameter from
the fundamental flavor degrees. In the light cone coordinates, the
metric (\ref{background}) becomes \bea
ds^2&=&\frac{R^2h(z)}{z^2}(-(1+f)dx^+dx^-+d{x_2}^2+d{x_3}^2\nonumber \\
&&+\frac{1}{2}(1-f)({dx^+}^2+{dx^-}^2)+ \frac{1}{f}dz^2),
\label{metric3}\eea  and the other parts are same as in the metric
(\ref{background}). Choosing the gauge $\tau=X^-, ~\sigma=X^2,
~z=z(\sigma)$, and substituting the induced metric of the
fundamental string into the Nambu-Goto action, we get \be
S=\frac{R^2L^-}{\sqrt{2}\pi\alpha'z_T^2}\int_0^{L/2} d\sigma
e^{\varepsilon z^2}\sqrt{1+f^{-1}z'^2}.\label{action}\ee And the
equation of motion of $z(\sigma)$ is \be z'^2=f(e^{2\varepsilon
z^2}-B), \ee where the $B$ is an integral constant. Choosing the
following boundary condition $z(\sigma=\pm\frac{L}{2})=0$, and
$z'(\sigma=0)=0$, then the turning point will be at $z=z_T$ or
$z=(\frac{\ln B}{2\varepsilon})^{1/2}$.

In the following, we consider the $z=z_T$ case. Since the parameter
$\varepsilon$ is very small, then the length of string in the
background (\ref{metric3}) to the leading order is \be
\frac{L}{2}=\int^{z_T}_0\frac{dz}{\sqrt{f(e^{2\varepsilon
z^2}-B)}}=\frac{az_T}{\gamma}-\varepsilon\frac{bz_T^3}{\gamma^3},\label{gamma}\ee
where the $\gamma^2=1-B$,
$a=\frac{\sqrt{\pi}\Gamma(\frac{5}{4})}{\Gamma(\frac{3}{4})}$ and
$b=\frac{\sqrt{\pi}\Gamma(\frac{3}{4})}{\Gamma(\frac{1}{4})}$. Using
the definition $R^2/\alpha'=\sqrt{\lambda}$, and the condition
$LT\ll 1$ of the wilson loop, then, to the leading order
$\varepsilon$, the difference between the action (\ref{action}) and
the self-energy of the fundamental string reads \be  S-S_0=
\frac{\pi^2}{8\sqrt{2}a}\sqrt{\lambda}L^2L^-T^3-\varepsilon\frac{b}{8\sqrt{2}a^2}
\sqrt{\lambda}L^2L^-T. \ee

By the nonperturbative definition of the jet-quenching parameter
$\hat{q}= \frac{8\sqrt{2}}{L^2L^-}(S-S_0)$ in \cite{Liu:2006ug}, the
jet-quenching parameter to the leading order is \be \hat{q}=
\frac{\pi^2}{a}\sqrt{\lambda}T^3-\varepsilon\frac{b}{a^2}\sqrt{\lambda}T.\label{q}\ee
Thus, to be compared with the $\mathcal{N}=4$ SYM,
$\hat{q}=$$\frac{\pi^2}{a}\sqrt{\lambda}T^3$, the jet-quenching
parameter in the QCD-like holographic model is decreased due to the
different degrees of freedom between the QCD-like model and the
$\mathcal{N}=4$ SYM.

In the above, we have fixed the parameter $\varepsilon=3cN_f/N_c$.
Thus, through substituting into the equation (\ref{q}), the
jet-quenching parameter can be rewritten as \be \hat{q}=
\frac{\pi^2}{a}\sqrt{\lambda}T^3-\frac{N_f}{N_c}
\frac{3bc}{a^2}\sqrt{\lambda}T.\label{jet}\ee Thus, we can see that
the last term comes from the contribution of the fundamental
flavors. And it is proportional to the ratio $N_f/N_c$. If $N_f=0$,
then it will reduce to the case of the $\mathcal{N}=4$ SYM QGP. The
temperature dependent of this correction is linear, so if increasing
the temperature, then the correction will become large. But, due to
the power three temperature dependence of the first term in the
equation (\ref{jet}), the full jet quenching parameter also becomes
large. Since this correction is negative, the jet quenching
parameter will be smaller than that of the thermal $\mathcal{N}=4$
SYM. So this result provides an evidence to support the conjecture
proposed in \cite{Liu:2006he}, which state the jet quenching
parameter $\hat{q}$ of nonconformal theories always decreases along
renormalization group trajectories. In our case, we can consider the
parameter $\varepsilon$ or $N_f/N_c$ as a varying scale. Then the
jet quenching parameter $\hat{q}$ can flow from the $\mathcal{N}=4$
to the QCD-like models by varying the parameter $\varepsilon$.

We can give a comparison with the RHIC data. Choosing the $N_c=3$,
$N_f=1$ and $\lambda=6\pi$ as same in \cite{Liu:2006ug}, then, for
the temperature $T=300$ \textrm{Mev}, $T=400$ \textrm{Mev} and
$T=500$ \textrm{Mev}, the first term in the equation (\ref{jet}) is
same as in the \cite{Liu:2006ug}. And the second term which comes
from the contributions of the fundamental flavors is $0.26$
\textrm{Gev$^2/$fm}, $0.34$ \textrm{Gev$^2/$fm} and $0.43$
\textrm{Gev$^2/$fm}. So the full jet quenching parameter is $4.24$
\textrm{Gev$^2/$fm}, $10.26$ \textrm{Gev$^2/$fm} and $20.27$
\textrm{Gev$^2/$fm}. From the RHIC data, the parameter is in the
range of $5$-$15$ \textrm{Gev$^2/$fm}. So our results except for the
case at the temperature $T=300$ \textrm{Mev} are in this range.

\noindent{\bf Conclusions:} From the AdS/QCD method, the holographic
models \cite{Karch:2006pv, DaRold:2005zs, Andreev:2006vy} can
produce many interesting properties of the QCD. The gravity
backgrounds of these models are AdS$_5$-Schwarz-like. We proposed
that the reason of deformation of the AdS$_5$-Schwarz geometry is
the different degrees of freedom of QCD relative to the
$\mathcal{N}=4$ SYM, mainly the fundamental flavors. Then the
parameter $\varepsilon$ describing the deformation, to leading
order, will be proportional to the $N_f/N_c$. So in this deformed
AdS$_5$-Schwarz background, we can calculate the contributions of
fundamental flavors to the shear viscosity and jet quenching
parameter.

For corrections to the shear viscosity $\eta$ and the ratio
$\eta/s$, the $N_f$ and $N_c$ dependence is all the same as that
obtained by \cite{Mateos:2006yd}. In the above, the temperature
dependence is linear for the shear viscosity, and $T^{-2}$ for the
ratio $\eta/s$. Obviously, they are different from that in
\cite{Mateos:2006yd}. After comparing with the RHIC data, we find
our results for the holographic QCD-like model are completely in the
range of experiment value. For the jet quenching parameter of the
thermal QGP, the correction from the fundamental flavors is
proportional to $N_f/N_c$ as the ratio $\eta/s$. While the
temperature dependence is linear, which is the same as that of shear
viscosity. Since this correction is negative, the jet quenching
parameter is smaller than that of the $\mathcal{N}=4$ SYM. Hence it
is an evidence supporting the conjecture proposed by
\cite{Liu:2006he}.

From our results for the ratio $\eta/s$ and the jet quenching
parameter $\hat{q}$, we see that the parameter $\varepsilon$ or
$N_f/N_c$ is something like a scale. By varying the parameter
$\varepsilon$ or $N_f/N_c$, the $\mathcal{N}=4$ SYM will flow to the
holographic QCD-like models.

\noindent{\bf Acknowledgments:} We would like to thank J.P. Shock
for some useful discussions.



\begin{thebibliography}{40}

\bibitem{Maldacena:1997re}
  J.~M.~Maldacena,
  Adv.\ Theor.\ Math.\ Phys.\  {\bf 2}, 231 (1998)
  [Int.\ J.\ Theor.\ Phys.\  {\bf 38}, 1113 (1999)]
  [arXiv:hep-th/9711200];
  S.~S.~Gubser, I.~R.~Klebanov and A.~M.~Polyakov,
  Phys.\ Lett.\  B {\bf 428}, 105 (1998)
  [arXiv:hep-th/9802109];
  E.~Witten,
  Adv.\ Theor.\ Math.\ Phys.\  {\bf 2}, 253 (1998)
  [arXiv:hep-th/9802150];
  E.~Witten,
  Adv.\ Theor.\ Math.\ Phys.\  {\bf 2}, 505 (1998)
  [arXiv:hep-th/9803131].
  O.~Aharony, S.~S.~Gubser, J.~M.~Maldacena, H.~Ooguri and Y.~Oz,
  Phys.\ Rept.\  {\bf 323}, 183 (2000)
  [arXiv:hep-th/9905111].

\bibitem{Baier:1996sk}
  R.~Baier, Y.~L.~Dokshitzer, A.~H.~Mueller, S.~Peigne and D.~Schiff,
  Nucl.\ Phys.\  B {\bf 484}, 265 (1997)
  [arXiv:hep-ph/9608322].

\bibitem{Policastro:2001yc}
  G.~Policastro, D.~T.~Son and A.~O.~Starinets,
  Phys.\ Rev.\ Lett.\  {\bf 87}, 081601 (2001)
  [arXiv:hep-th/0104066].

\bibitem{Kovtun:2004de}
  P.~Kovtun, D.~T.~Son and A.~O.~Starinets,
  Phys.\ Rev.\ Lett.\  {\bf 94}, 111601 (2005)
  [arXiv:hep-th/0405231].

\bibitem{Kovtun:2003wp}
  P.~Kovtun, D.~T.~Son and A.~O.~Starinets,
  JHEP {\bf 0310} (2003) 064
  [arXiv:hep-th/0309213].

\bibitem{Policastro:2002se}
  D.~T.~Son and A.~O.~Starinets,
  JHEP {\bf 0209}, 042 (2002)
  [arXiv:hep-th/0205051];
  G.~Policastro, D.~T.~Son and A.~O.~Starinets,
  JHEP {\bf 0209}, 043 (2002)
  [arXiv:hep-th/0205052].

\bibitem{Buchel:2003tz}
  A.~Buchel and J.~T.~Liu,
  Phys.\ Rev.\ Lett.\  {\bf 93}, 090602 (2004)
  [arXiv:hep-th/0311175].

\bibitem{Buchel:2004di}
  A.~Buchel, J.~T.~Liu and A.~O.~Starinets,
  Nucl.\ Phys.\  B {\bf 707} (2005) 56
  [arXiv:hep-th/0406264].


\bibitem{Mateos:2006yd}
  D.~Mateos, R.~C.~Myers and R.~M.~Thomson,
  Phys.\ Rev.\ Lett.\  {\bf 98}, 101601 (2007)
  [arXiv: hep-th/0610184].

\bibitem{Liu:2006ug}
  H.~Liu, K.~Rajagopal and U.~A.~Wiedemann,
  Phys.\ Rev.\ Lett.\  {\bf 97}, 182301 (2006)
  [arXiv:hep-ph/0605178].

\bibitem{Herzog:2006gh}
  C.~P.~Herzog, A.~Karch, P.~Kovtun, C.~Kozcaz and L.~G.~Yaffe,
  JHEP {\bf 0607}, 013 (2006)
  [arXiv:hep-th/0605158];
  S.~S.~Gubser,
  Phys.\ Rev.\  D {\bf 74}, 126005 (2006)
  [arXiv:hep-th/0605182];
  J.~Casalderrey-Solana and D.~Teaney,
  Phys.\ Rev.\  D {\bf 74}, 085012 (2006)
  [arXiv:hep-ph/0605199].

\bibitem{Buchel:2006bv}
  A.~Buchel,
  Phys.\ Rev.\  D {\bf 74}, 046006 (2006)
  [arXiv:hep-th/0605178].

\bibitem{Armesto:2006zv}
  N.~Armesto, J.~D.~Edelstein and J.~Mas,
  JHEP {\bf 0609}, 039 (2006)
  [arXiv:hep-ph/0606245].

\bibitem{Vazquez-Poritz:2006ba}
  J.~F.~Vazquez-Poritz,
  arXiv:hep-th/0605296.

\bibitem{Gubser:2006nz}
  S.~S.~Gubser,
  arXiv:hep-th/0612143.

\bibitem{Argyres:2006yz}
  P.~C.~Argyres, M.~Edalati and J.~F.~Vazquez-Poritz,
  JHEP {\bf 0704}, 049 (2007)
  [arXiv:hep-th/0612157].

\bibitem{Gao:2006uf}
  Y.~h.~Gao, W.~s.~Xu and D.~f.~Zeng,
  arXiv:hep-th/0611217.

\bibitem{Liu:2006he}
  H.~Liu, K.~Rajagopal and U.~A.~Wiedemann,
  JHEP {\bf 0703}, 066 (2007)
  [arXiv:hep-ph/0612168].

\bibitem{Bertoldi:2007sf}
  G.~Bertoldi, F.~Bigazzi, A.~L.~Cotrone and J.~D.~Edelstein,
  arXiv:hep-th/0702225.

\bibitem{Casero:2006pt}
  R.~Casero, C.~Nunez and A.~Paredes,
  Phys.\ Rev.\  D {\bf 73}, 086005 (2006)
  [arXiv:hep-th/0602027]; F.~Benini, F.~Canoura, S.~Cremonesi, C.~Nunez and A.~V.~Ramallo,
  JHEP {\bf 0702}, 090 (2007)
  [arXiv:hep-th/0612118].

\bibitem{Cotrone:2007qa}
  A.~L.~Cotrone, J.~M.~Pons and P.~Talavera,
  arXiv:0706.2766 [hep-th].

\bibitem{Karch:2006pv}
   A.~Karch, E.~Katz, D.~T.~Son and M.~A.~Stephanov,
  Phys.\ Rev.\ Lett.\  {\bf 95}, 261602 (2005)
  [arXiv:hep-ph/0501128];
  Phys.\ Rev.\  D {\bf 74}, 015005 (2006)
  [arXiv:hep-ph/0602229].

\bibitem{DaRold:2005zs}
  L.~Da Rold and A.~Pomarol,
  Nucl.\ Phys.\  B {\bf 721}, 79 (2005)
  [arXiv:hep-ph/0501218].

\bibitem{Andreev:2006vy}
  O.~Andreev,
  Phys.\ Rev.\  D {\bf 73}, 107901 (2006)
  [arXiv:hep-th/0603170];
  O.~Andreev and V.~I.~Zakharov,
  Phys.\ Rev.\  D {\bf 74}, 025023 (2006)
  [arXiv:hep-ph/0604204];
  Phys.\ lett.\  B {\bf 645}, 437 (2007)
  [arXiv:hep-ph/0607026].

\bibitem{Herzog:2006ra}
  C.~P.~Herzog,
  Phys.\ Rev.\ Lett.\  {\bf 98}, 091601 (2007)
  [arXiv:hep-th/0608151].

\bibitem{Nakano:2006js}
  E.~Nakano, S.~Teraguchi and W.~Y.~Wen,
  Phys.\ Rev.\  D {\bf 75}, 085016 (2007)
  [arXiv:hep-ph/0608274].

\end{thebibliography}
\end{document}